\documentclass[prd,superscriptaddress,amsfonts,amssymb,amsmath,showpacs]{revtex4}

\usepackage{amsmath}
\usepackage{amssymb}

\usepackage{braket}
\usepackage{eurosym}
\usepackage{amsfonts}
\usepackage{graphicx}
\usepackage{calrsfs}
\usepackage[usenames,dvipsnames,svgnames,table]{xcolor}
\usepackage{hyperref}

\begin{document}

\title{Stable Dyonic Thin-Shell Wormholes in Low-Energy String Theory }

\author{Ali \"{O}vg\"{u}n}
\email{ali.ovgun@pucv.cl}

\affiliation{Instituto de F\'{\i}sica, Pontificia Universidad Cat\'olica de
Valpara\'{\i}so, Casilla 4950, Valpara\'{\i}so, Chile}

\affiliation{Physics Department, Arts and Sciences Faculty, Eastern Mediterranean University, Famagusta, North Cyprus via Mersin 10, Turkey}

\author{Kimet Jusufi}
\email{kimet.jusufi@unite.edu.mk}

\affiliation{Physics Department, State University of Tetovo, Ilinden Street nn,
1200, Macedonia}
\affiliation{
Institute of Physics, Faculty of Natural Sciences and Mathematics, Ss. Cyril and Methodius University, Arhimedova 3, 1000 Skopje, Macedonia}

\date{\today }

\begin{abstract}
Considerable attention has been devoted to the wormhole physics in the past 30 years by exploring the possibilities of finding traversable wormholes without the need of exotic matter. In particular the thin-shell wormhole formalism has been widely investigated by exploiting the cut-and-paste technique to merge two space-time regions and, to research the stability of these wormholes developed by Visser. This method helps us to minimize the amount of the exotic matter. In this paper we construct a four dimensional, spherically symmetric, dyonic thin-shell wormhole with electric charge $Q$, magnetic charge $P$, and dilaton charge $\Sigma$, in the context of Einstein-Maxwell-dilaton theory.  We have applied  Darmois-Israel formalism and the cut-and-paste method by joining together two identical spacetime solutions. We carry out the dyonic thin-shell wormhole stability analyses by using a linear barotropic gas, Chaplygin gas, and logarithmic gas for the exotic matter. It is shown that by choosing suitable parameter values as well as equation of state parameter, under specific conditions we obtain a stable dyonic thin-shell wormhole solution. Finally we argue that, the stability domain of the dyonic thin-shell wormhole can be increased in terms of electric charge, magnetic charge, and dilaton charge. 
\end{abstract}


\maketitle
 
\section{Introduction}

Wormholes are exotic objects predicted by Einstein's theory of gravity which act as a space-time tunnel by connecting two different regions of the universe. Thought, the idea of wormholes is not new \cite{fl,er}, the interest on wormholes was recently reborn by the seminal work of Morris and Thorne \cite{mth1} who studied the traversable wormholes. There are, however, several problematic issues related to the possible existence of wormholes, in particular, it was shown that the existence of wormholes require the violation of energy conditions \cite{mth2,ao}. Another major problem is related to the stability analysis of wormholes. On the other hand, Visser attempted to minimize the existence of the exotic matter by constructing infinitesimally small thin-shell wormholes \cite{mv1,mv2,mv3,Poisson:1995sv}. The Visser's method is based on the cut-and-paste technique by joining together two identical space-time solutions and making use of the Darmois-Israel formalism \cite{is} to compute the surface stress-energy tensor components. Finally these results can be used to study the wormhole dynamics with the help of Lanczos equations.

This method was applied to construct a number of thin-shell wormholes (TSW), including charged TSW \cite{eiroa1,ayan1}, TSW with a cosmological constant \cite{lobo1}, TSW in dilaton gravity \cite{eiroa2}, TSW from the regular Hayward black hole \cite{ali1}, TSW in higher dimensional Einstein-Maxwell theory \cite{rah,rahh}, rotating TSW \cite{mazh1,ali}, quantum corrected TSW in Bohmian quantum mechanics \cite{kimet},  primordial wormholes induced from a Grand Unified Theories (GUTs) \cite{odintsov1,odintsov2}, canonical acoustic TSW, charged TSW with dilaton field, TSW with a Chaplygin gas, traversable wormholes in the anti-de Sitter spacetime, TSW with a negative cosmological constant, wormholes in mimetic gravity, TSW from charged black string, cylindrical TSW, and many other interesting papers \cite{23,24,25,26,27,28,29,30,31,32,33,34,35,36,37,38,39,40,41,42,43,44,45,46,47,48,49,50,51,52,53,54,55,56,57,58,59,60}. While the stability analyses is investigated by different models, for example, linear perturbations \cite{s1} and specific equations of state (EoS) such as: linear barotropic gas (LBG), Chaplygin gas (CG), and logarithmic gas (LogG) for the exotic matter \cite{s2,s3,s4,s5,s6}. 

Recently Goulart found a four-dimensional, spherically symmetric, dyonic black hole and charged wormhole solution in the low-energy effective actions of string theory or supergravity theory \cite{Goulart,Goulart1}. Furthermore in \cite{pedro}, a time-dependent spherically symmetric black hole solution in the context of low-energy string theory was investigated. The solution was found by Goulart, is of particular interest since it can be written in terms of five independent parameters: the electric charge $Q$, the magnetic charge $\Sigma$, the value of the dilation of infinity $\phi_0$, and two integration constants, $r_ {1} $ and $r_ {2} $. Inspired by this work, we aim to use this solution and construct a four-dimensional TSW wormhole in the context of Einstein-Maxwell-dilaton (EMD) theory and then investigate the role of electric charge $Q$, magnetic charge $P$, and dilaton charge $\Sigma$, on the stability domain of the wormhole.

The structure of this paper is as follows: in Section II, we review briefly the dyonic black hole solutions. In Section III, using
the Visser's cut-and-paste technique we construct dyonic thin-shell wormhole (DTSW). In Section IV, we check the stability conditions for different types of
gases such as a LBG, CG, and LogG for the exotic matter. In section V, we comment on our results.

\section{Dyonic Black Holes in the EMD theory}

In this part we use the dyonic black hole solutions in the EMD theory found by Goulart \cite{Goulart}. Firstly, we
consider the action of the EMD without a dilaton
potential and without an axion 
\begin{equation}
S=\int d^{4}x\sqrt{-g}\left(R-2\partial_{\mu}\phi\partial^{\mu}\phi-W(\phi)F_{\mu\nu}F^{\mu\nu}\right).\label{1}
\end{equation}
where the field strength is given
by 
\begin{equation}
F_{\mu\nu}=\partial_{\mu}A_{\nu}-\partial_{\nu}A_{\mu}.
\end{equation}
Furthermore, for constant axion field the bosonic sector of $SU(4)$
version of $\mathcal{N}=4$ supergravity theory is $W(\phi)=e^{-2\phi}$
\cite{41}. It is noted that there are five independent parameters
such as $Q,P,\phi_{0},r_{1}$ and $r_{2}$. Accordingly, the spacetime
of the general spherically symmetric solution is given by the line
element \cite{Goulart} 
\begin{equation}
ds^{2}=-f(r)dt^{2}+\frac{1}{f(r)}dr^{2}+h(r)(d\theta^{2}+\sin^{2}\theta d\varphi^{2}),\label{3}
\end{equation}
where 
\begin{align}
f(r) & =\frac{(r-r_{1})(r-r_{2})}{(r+d_{0})(r+d_{1})},\,\,\,h(r)=(r+d_{0})(r+d_{1}),\label{4}\\
e^{2\phi} & =e^{2\phi_{0}}\frac{r+d_{1}}{r+d_{0}},\label{5}\\
F_{rt} & =\frac{e^{2\phi_{0}}Q}{(r+d_{0})^{2}},\,\,\,F_{\theta\varphi}=P\sin\theta,\label{6}
\end{align}
with 
\begin{align}
d_{0} & =\frac{-(r_{1}+r_{2})\pm\sqrt{(r_{1}-r_{2})^{2}+8e^{2\phi_{0}}Q^{2}}}{2},\label{7}\\
d_{1} & =\frac{-(r_{1}+r_{2})\pm\sqrt{(r_{1}-r_{2})^{2}+8e^{-2\phi_{0}}P^{2}}}{2}.\label{8}
\end{align}
Note that the corresponding electric and magnetic charges are $Q$
and $P$, respectively. The $\phi_{0}$ stands for the value of the
dilaton at infinity. Furthermore, there are two integration constants
such as $r_{1}$ and $r_{2}$. On the other hand, $d_{0}$ and $d_{1}$
are dependent constants inasmuch as they transform into each other
under S-duality, i.e. $Q\leftrightarrow P$ and $\phi\rightarrow-\phi$. It is noted that $e^{-2\phi}$, which is the the dilaton
coupling, is also invariant. For the extremal limit, we choose their
signs to be same to avoid negative entropies and without loss of generality
it is taken as $r_{2}>r_{1}$. Here $r_{1}$ and $r_{2}$ are the
inner and outer horizons respectively \cite{Goulart}.

The Hawking temperature is calculated by following 
\begin{equation}
T=\frac{1}{4\pi}\frac{(r_{2}-r_{1})}{(r_{2}+d_{0})(r_{2}+d_{1})},\label{9}
\end{equation}
and the entropy of the black hole is 
\begin{equation}
S=\pi(r_{2}+d_{0})(r_{2}+d_{1}).
\end{equation}
One can also define the dilaton charge as follows 
\begin{equation}
\Sigma=\frac{1}{4\pi}\int d\Sigma^{\mu}\nabla_{\mu}\phi=\frac{(d_{0}-d_{1})}{2},
\end{equation}
 depending on the values of electric/magnetic charge of black hole,
it can be positive or negative. Firstly, the four parameters ($Q,P,\phi_{0},
M)$ dyonic solution is found in \cite{37}. Here the key point is
that there is no boundary condition on the $r_{1}$ and $r_{2}$ to
make this dyonic black hole. 

The Ricci scalar is calculated as follows
\begin{equation}
R=\frac{(d_{0}-d_{1})^{2}(r-r_{1})(r-r_{2})}{2(r+d_{0})^{3}(r+d_{1})^{3}}.
\end{equation}
The domain of the $h(r)\geq0$ is restricted with the causality. The
singularity is found at $r_{S}=-d_{0}$ for $d_{0}>d_{1}$, or at
$r_{S}=-d_{1}$ for $d_{1}>d_{0}$.

One of the special case which we use to construct a DTSW
when $d_{1}=-d_{0}$ that the dilaton charge is a constant $d_{0}$
such as $d_{0}=\Sigma$. Furthermore we suppose that $(r_{1}+r_{2})=2M$
and ($r_{1}r_{2}$)= $r_{0}^{2}$ \cite{37}. The solution becomes
\begin{align}
f(r) & =\frac{(r-r_{1})(r-r_{2})}{(r^{2}-\Sigma^{2})},\,\,h(r)=(r^{2}-\Sigma^{2})\label{metKal}\\
e^{2\phi} & =e^{2\phi_{0}}\frac{r-\Sigma}{r+\Sigma},\\
F_{rt} & =\frac{e^{2\phi_{0}}Q}{(r+\Sigma)^{2}},\,\,\,F_{\theta\varphi}=P\sin\theta.
\end{align}
 One can find the magnetically charged solutions of \cite{38} and
\cite{39} by using $Q=0$ and also the Schwarzschild solution by
setting that $P=0$.

\section{Construction of DTSW}

Let us now proceed to use cut-and-paste technique to construct a DTSW using the
metric (3). Consider two spherically symmetric spacetime solutions of the dyonic black hole metric in 4-dimensions, then remove from each four-dimensional manifold $M^{(\pm)}$ the regions described by \cite{Poisson:1995sv}
\begin{equation}
M^{(\pm)}=\left\lbrace r^{(\pm)}\leq a|a>r_{h}\right\rbrace ,
\end{equation}
where $a$ is the radius of the throat of the DTSW with an important condition $a>r_{h}$. In other words, $a$ should be greater than the event horizon in order to avoid the formation of an event horizon. Next, paste these two manifolds at the boundary hypersurface given by $\Sigma^{(\pm)}=\left\lbrace r^{(\pm)}=a,a>r_{h}\right\rbrace $ which results with 
a geodesically complete manifold $M=M^{+}\bigcup M^{-}$.
According to the Darmois-Israel formalism we can choose the coordinates on $M$ as $x^{\alpha}=(t,r,\theta,\varphi)$, while the coordinates on the induced metric $\Sigma$ are $\xi^{i}=(\tau,\theta,\varphi)$.
For the parametric equation on $\Sigma$ we can write
\begin{equation}
\Sigma:F(r,\tau)=r-a(\tau)=0.
\end{equation}

Our main goal is to compare various characteristics of EMD theory and dyonic black hole, such as the surface stress-energy tensor and the basic question of stability. For this purpose, we define the  dynamical induced metric on $\Sigma$ that can now be written in terms of the the proper time
$\tau$ on the shell, where $a=a(\tau)$, as follows
\begin{equation}
\mathrm{d}s_{\Sigma}^{2}=-\mathrm{d}\tau^{2}+a(\tau)^{2}\left(\mathrm{d}\theta^{2}+\sin^{2}\theta\,\mathrm{d}\varphi^{2}\right).
\end{equation}

The junction conditions on $\Sigma$ implies from the Lanczos equations
\begin{equation}
{S^{i}}_{j}=-\frac{1}{8\pi}\left(\left[{K^{i}}_{j}\right]-{\delta^{i}}_{j}\,K\right),
\end{equation}
in which ${S^{i}}_{j}=diag(-\sigma,p_{\theta},p_{\varphi})$
is the energy momentum tensor on the thin-shell, $K$ and $[K_{ij}]$,
are defined as $K=trace\,[{K^{i}}_{i}]$ and $[K_{ij}]={K_{ij}}^{+}-{K_{ij}}^{-}$,
respectively. Furthermore the extrinsic curvature ${K^{i}}_{j}$ is defined
by 
\begin{equation}
K_{ij}^{(\pm)}=-n_{\mu}^{(\pm)}\left(\frac{\partial^{2}x^{\mu}}{\partial\xi^{i}\partial\xi^{j}}+\Gamma_{\alpha\beta}^{\mu}\frac{\partial x^{\alpha}}{\partial\xi^{i}}\frac{\partial x^{\beta}}{\partial\xi^{j}}\right)_{\Sigma}.
\end{equation}

We can choose the unit vectors  ${n_{\mu}}^{(\pm)}$, such that $n_{\mu}n^{\mu}=1$ and normal
to $M^{(\pm)}$ as follows
\begin{equation}
n_{\mu}^{(\pm)}=\pm\left(\left\vert g^{\alpha\beta}\frac{\partial F}{\partial x^{\alpha}}\frac{\partial F}{\partial x^{\beta}}\right\vert ^{-1/2}\frac{\partial F}{\partial x^{\mu}}\right)_{\Sigma}.
\end{equation}

Adopting the orthonormal basis $\{e_{\hat{\tau}},e_{\hat{\theta}},e_{\hat{\varphi}}\}$
($e_{\hat{\tau}}=e_{\tau}$, $e_{\hat{\theta}}=[h(a)]^{-1/2}e_{\theta}$,
$e_{\hat{\varphi}}=[h(a)\sin^{2}\theta]^{-1/2}e_{\varphi}$), for
the metric (3), the extrinsic curvature components are found
as \cite{24}
\begin{equation}
K_{\hat{\theta}\hat{\theta}}^{\pm}=K_{\hat{\varphi}\hat{\varphi}}^{\pm}=\pm\frac{h'(a)}{2h(a)}\sqrt{f(a)+\dot{a}^{2}},\label{eq5}
\end{equation}
and 
\begin{equation}
K_{\hat{\tau}\hat{\tau}}^{\pm}=\mp\frac{2\ddot{a}+f'(a)}{2\sqrt{f(a)+\dot{a}^{2}}},\label{eq6}
\end{equation}
where the prime and the dot represent the derivatives with respect
to $r$ and $\tau$, respectively. With the definitions of $[K_{_{\hat{\imath}\hat{\jmath}}}]\equiv K_{_{\hat{\imath}\hat{\jmath}}}^{+}-K_{_{\hat{\imath}\hat{\jmath}}}^{-}$,
and $K=tr[K_{\hat{\imath}\hat{\jmath}}]=[K_{\;\hat{\imath}}^{\hat{\imath}}]$,
and the introduction of the surface stress-energy tensor $S_{_{\hat{\imath}\hat{\jmath}}}={\rm diag}(\sigma,p_{\hat{\theta}},p_{\hat{\varphi}})$
we have the Einstein equations on the shell (also called the Lanczos
equations): 
\begin{equation}
-[K_{\hat{\imath}\hat{\jmath}}]+Kg_{\hat{\imath}\hat{\jmath}}=8\pi S_{\hat{\imath}\hat{\jmath}},\label{eq7}
\end{equation}
that in our case results in a shell of radius $a$ with energy density
$\sigma$ and transverse pressure $p=p_{\hat{\theta}}=p_{\hat{\varphi}}$.
Using the above results from the Lanczos equations, one can easily check that the surface density
and the surface pressure are given by the following relations \cite{24,25}
\begin{equation}
\sigma=-\frac{\sqrt{f(a)+\dot{a}^{2}}}{4\pi}\frac{h'(a)}{h(a)},\label{25}
\end{equation}
\begin{equation}
p=\frac{\sqrt{f(a)+\dot{a}^{2}}}{8\pi}\left[\frac{2\ddot{a}+f'(a)}{f(a)+\dot{a}^{2}}+\frac{h'(a)}{h(a)}\right].\label{26}
\end{equation}

Note that the energy density is negative at the throat because of the flare-out condition which the area is minimal at the throat (then $h(r)$ increases for $r$ close to $a$ and $h'(a)>0$) so we have exotic matter. From the last two equations we can now write the static configuration
of radius $a$, by setting $\dot{a}=0$, and $\ddot{a}=0$, we get 

\begin{equation}
\sigma_{0}=-\frac{\sqrt{f(a_{0})}}{4\pi}\frac{h'(a_{0})}{h(a_{0})},\label{27}
\end{equation}
and 
\begin{equation}
p_{0}=\frac{\sqrt{f(a_{0})}}{8\pi}\left[\frac{f'(a_{0})}{f(a_{0})}+\frac{h'(a_{0})}{h(a_{0})}\right].\label{28}
\end{equation}

From Eq. \eqref{27} we see that the surface density is negative, i.e. $\sigma_{0}<0$,
as a consequence of this, the (WEC) is violated. The amount of exotic matter concentrated at the wormhole is calculated by the following integral 
\begin{equation}
\Omega_{\sigma}=\int\sqrt{-g}\,\left(\rho+p_{r}\right)\,\mathrm{d}^{3}x.
\end{equation}

In the case of a TSW we have $p_{r}=0$ and $\rho=\sigma\delta(r-a)$,
where $\delta(r-a)$ is the Dirac delta function. The above integral can be easily evaluated if we first make use of the Dirac delta function 
\begin{equation}
\Omega_{\sigma}=\int_{0}^{2\pi}\int_{0}^{\pi}\int_{-\infty}^{\infty}\sigma\sqrt{-g}\,\delta(r-a)\mathrm{d}r\,\mathrm{d}\theta\,\mathrm{d}\varphi.
\end{equation}

Substituting the value of energy density in the last equation, for the energy density located on a thin-shell surface in static configuration we find
\begin{equation}
\Omega_{\sigma}=-2a_{0}\sqrt{\frac{(a_{0}-r_{1})(a_{0}-r_{2})}{(a_0^{2}-\Sigma^{2})}}.
\end{equation}

To analyze the attractive and repulsive nature of the wormhole we can calculate the 
observer's four-acceleration $a^{\mu}=u^{\nu}\nabla_{\nu}u^{\mu}$,
where the four velocity reads $u^{\mu}=(1/\sqrt{f(r)},0,0,0)$. For the radial 
component of the four-acceleration we find
\begin{equation}
a^{r}=\Gamma_{tt}^{r}\left(\frac{\mathrm{d}t}{\mathrm{d}\tau}\right)^{2}=\frac{a_0^{2}(r_{1}+r_{2})-2a_{0}(\Sigma^{2}+r_{1}r_{2})+\Sigma^{2}(r_{1}+r_{2})}{2(a_{0}^{2}-\Sigma^{2})^{2}}.
\end{equation}

One can easily observe that the test particle obeys the equation
of motion 
\begin{equation}
\frac{\mathrm{d}^{2}r}{\mathrm{d}\tau^{2}}=-\Gamma_{tt}^{r}\left(\frac{\mathrm{d}t}{\mathrm{d}\tau}\right)^{2}=-a^{r}.
\end{equation}

We conclude from the last equation that if  $a^{r}=0$, we get the geodesic equation, while the wormhole is attractive when $a^{r}>0$ and repulsive when a $a^{r}<0$.

\section{Stability Analysis}

In this Section, using the formalism developed in the Section (III), we calculate the potential and define the stability method for DTSW.  From the energy conservation
we have \cite{24}
\begin{equation}
\frac{d}{d\tau}\left(\sigma\mathcal{A}\right)+p\frac{d\mathcal{A}}{d\tau}=\left\lbrace \left[h'(a)\right]^{2}-2h(a)h''(a)\right\rbrace \frac{\dot{a}\sqrt{f(a)+\dot{a}^{2}}}{2h(a)},\label{34}
\end{equation}
where  the area of the wormhole is calculated by $\mathcal{A}=4\pi h(a)$. It is noted that the internal energy of the throat is located at the left side of Eq. \eqref{34} as a first term. Then the second term represents the work done
by the internal forces of the throat, on the other hand there is a flux term in the right side of the equation. Furthermore, to calculate the equation of dynamics of the wormhole, we use $\sigma(a)$ in Eq. \eqref{25} and find this simple equation,
\begin{equation}
\dot{a}^{2}=-V(a),\label{35}
\end{equation}
with potential
\begin{equation}
V(a)=f(a)-16\pi^{2}\left[\frac{h(a)}{h'(a)}\sigma(a)\right]^{2}.\label{36}
\end{equation}
A Taylor expansion to second order of the potential $V(a)$ around
the static solution yields \cite{24}: 
\begin{equation}
V(a)=V(a_{0})+V^{\prime}(a_{0})(a-a_{0})+\frac{V^{\prime\prime}(a_{0})}{2}(a-a_{0})^{2}+O(a-a_{0})^{3}.\label{37}
\end{equation}
From Eq. (\ref{36}) the first derivative of $V(a)$ is 
\begin{equation}
V^{\prime}(a)=f^{\prime}(a)-32\pi^{2}\sigma(a)\frac{h(a)}{h'(a)}\left\lbrace \left[1-\frac{h(a)h''(a)}{[h'(a)]^{2}}\right]\sigma(a)+\frac{h(a)}{h'(a)}\sigma^{\prime}(a)\right\rbrace ,\label{38}
\end{equation}
the last equation takes the form 
\begin{equation}
V^{\prime}(a)=f^{\prime}(a)+16\pi^{2}\sigma(a)\frac{h(a)}{h'(a)}\left[\sigma(a)+2p(a)\right].\label{39}
\end{equation}
The second derivative of the potential is 
\begin{eqnarray}
V^{\prime\prime}(a) & = & f^{\prime\prime}(a)+16\pi^{2}\left\lbrace \left[\frac{h(a)}{h'(a)}\sigma^{\prime}(a)+\left(1-\frac{h(a)h''(a)}{[h'(a)]^{2}}\right)\sigma(a)\right]\left[\sigma(a)+2p(a)\right]\right.\\
 &  & \left.+\frac{h(a)}{h'(a)}\sigma(a)\left[\sigma^{\prime}(a)+2p^{\prime}(a)\right]\right\rbrace .\nonumber \label{40}
\end{eqnarray}

Since $\sigma^{\prime}(a)+2p^{\prime}(a)=\sigma^{\prime}(a)[1+2p^{\prime}(a)/\sigma^{\prime}(a)]$,
replacing the parameter $p=\psi(\sigma)$ and $\psi'=dp/d\sigma=p^{\prime}/\sigma^{\prime}$,
we have that $\sigma^{\prime}(a)+2p^{\prime}(a)=\sigma^{\prime}(a)(1+2\psi')$,
and using Eq. \eqref{40} again, we obtain 
\begin{equation}
V^{\prime\prime}(a_{0})=f^{\prime\prime}(a_{0})-8\pi^{2}\left\lbrace \left[\sigma_{0}+2p_{0}\right]^{2}+2\sigma_{0}\left[\left(\frac{3}{2}-\frac{h(a_{0})h''(a_{0})}{[h'(a_{0})]^{2}}\right)\sigma_{0}+p_{0}\right](1+2\psi')\right\rbrace .\label{p10}
\end{equation}

The wormhole is stable if and only if $V^{\prime\prime}(a_{0})>0$.

\subsection{Stability analysis of DTSW via the LBG}

In what follows, we will use three different gas models for the exotic matter to explore the stability analysis; a LBG \cite{40}, CG \cite{s5,s6}, and
finally LogG \cite{ali1}. 

The equation of state of LBG \cite{s1,s2,s3,s4} is given by 
\begin{equation}
\psi=\omega\sigma,
\end{equation}
and hence 
\begin{equation}
\psi^{\prime}(\sigma_{0})=\omega,
\end{equation}
where $\omega$ is a constant parameter. For more useful information as
regards the effects of the parameters $\Sigma$, $r_{1}$ and $r_{2}$, we show graphically the DTSW stability
in terms of $\omega$ and $a_{0}$, as depicted in Fig. 1. 

By changing the values of $\Sigma$, $r_{1}$ and $r_{2}$, which encodes the effects of electric $Q$, magnetic $P$, and dilaton charge $\phi_{0}$ on the DTSW stability, we see from the Fig.1 that in two cases the region of stability is below the curve in the interval to the right of the asymptote, while in two other cases the stability region is simply below the curve. The region of stability is denoted by S.

\begin{figure}[h!]
\includegraphics[width=0.33\textwidth]{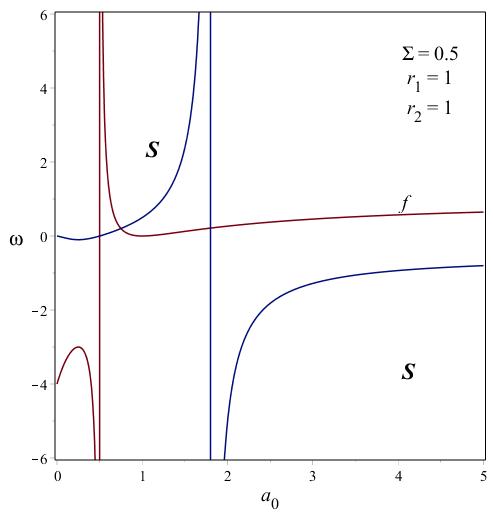} %
\includegraphics[width=0.33\textwidth]{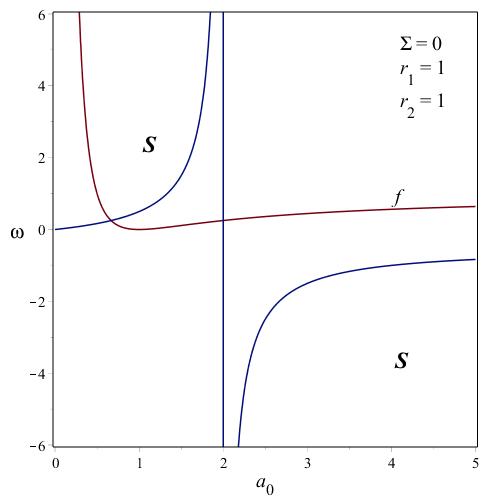} %
\includegraphics[width=0.33\textwidth]{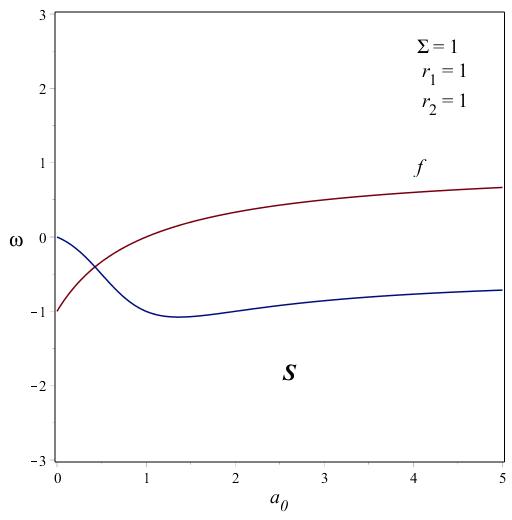} %
\includegraphics[width=0.33\textwidth]{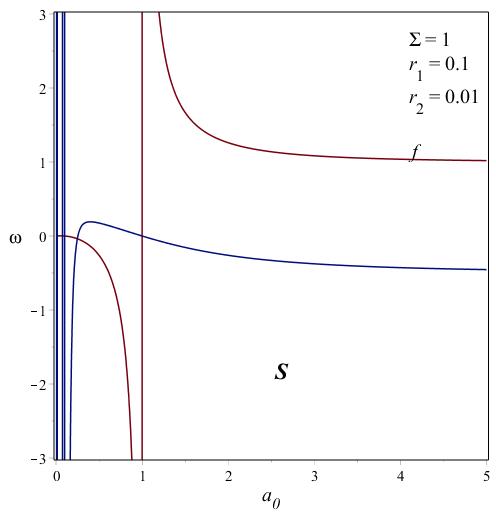}
\caption{\small \textit{Stability regions of DTSW  in terms of $\omega$ and radius of the throat $a_{0}$ for  different values of $\Sigma$, $r_{1}$ and $r_{2}$.} }
\end{figure}

\bigskip
\subsection{Stability analysis of DTSW via CG}

The equation of state of CG that we considered is given by \cite{s5}
\begin{equation}
\psi=\omega(\frac{1}{\sigma}-\frac{1}{\sigma_{0}})+p_{0},\label{44}
\end{equation}

and one naturally finds 
\begin{equation}
\psi^{\prime}(\sigma_{0})=-\frac{\omega}{\sigma_{0}^{2}}.
\end{equation}

After inserting Eq. \eqref{44} into Eq. \eqref{35}, we plot the stability regions of DTSW supported by CG in terms of $V^{\prime\prime}(a_{0})$ and $a_{0}$ as shown in Fig. 2. It is  worth of mentioning that in three cases the region of stability is above the curve in the interval to the right of the asymptote, while in one case stability region is below the curve in the interval to the right of the asymptote. The region of stability is denoted by S.

\begin{figure}[h!]
\includegraphics[width=0.34\textwidth]{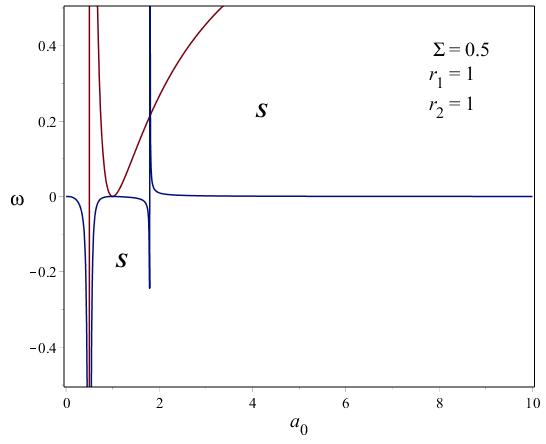} %
\includegraphics[width=0.33\textwidth]{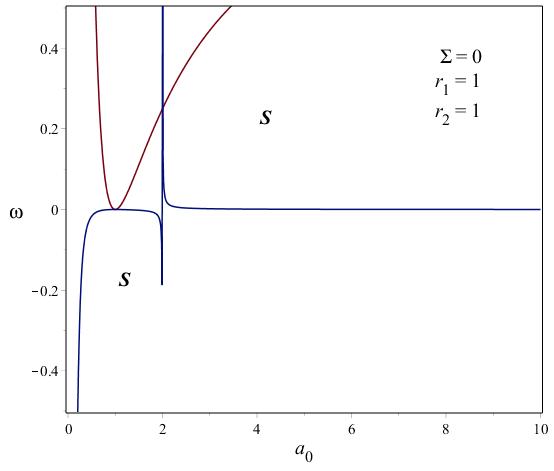} %
\includegraphics[width=0.34\textwidth]{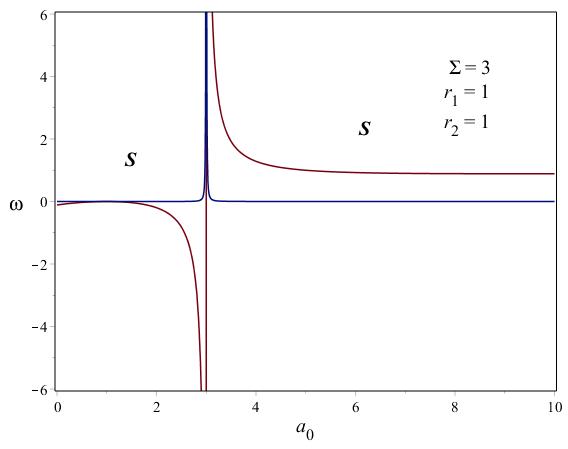} %
\includegraphics[width=0.33\textwidth]{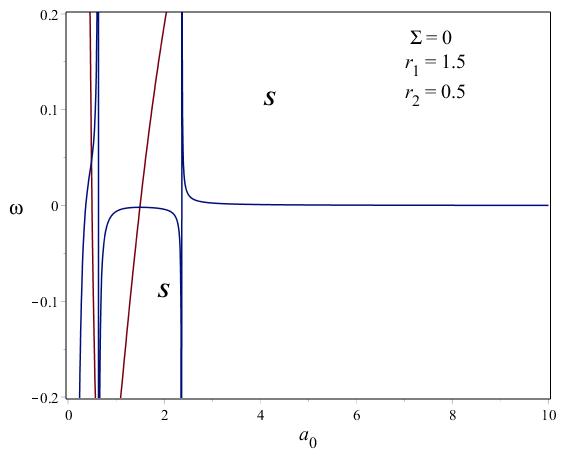} %
\caption{\small \textit{Stability regions of DTSW  in terms of $\omega$ as a function of the throat $a_{0}$ for  different values of $\Sigma$ and $r_{1}$, $r_{2}$.} }
\end{figure}

\bigskip
\subsection{Stability analysis of DTSW via LogG}

In our final example, the equation of state for LogG is selected as
follows \cite{ali1}
\begin{equation}
\psi=\omega\ln(\frac{\sigma}{\sigma_{0}})+p_{0},
\end{equation}
which leads to 
\begin{equation}
\psi^{\prime}(\sigma_{0})=\frac{\omega}{\sigma_{0}}.
\end{equation}
After inserting the above expression into Eq. \eqref{35}, we show the stability
regions of TSW supported by LogG in Fig. 3. In this case, we see that the region of stability is above the curve in the interval to the right of the asymptote. The region of stability is denoted by S. 

\begin{figure}[h!]
\includegraphics[width=0.30\textwidth]{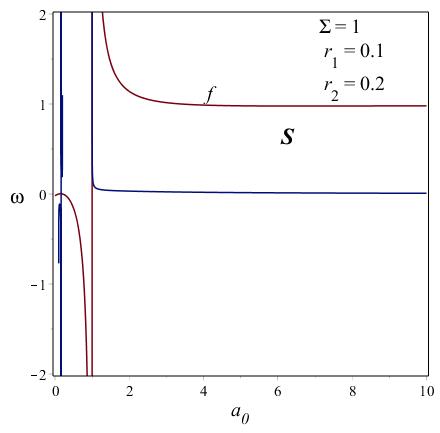} %
\includegraphics[width=0.30\textwidth]{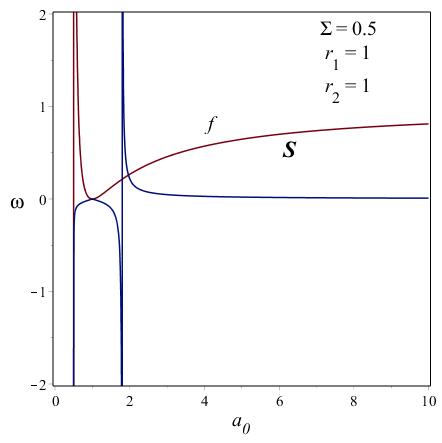} %
\includegraphics[width=0.30\textwidth]{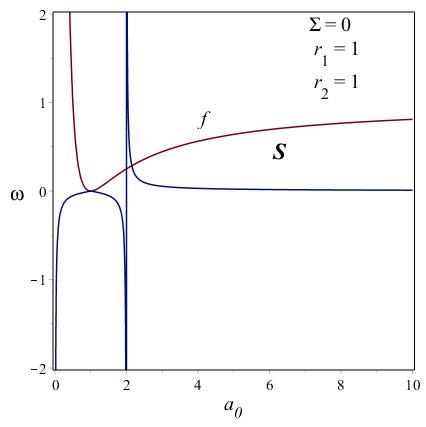} %
\includegraphics[width=0.30\textwidth]{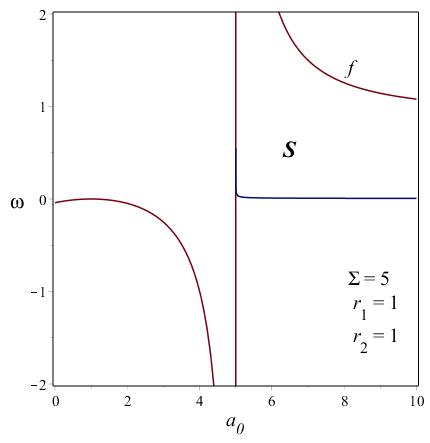} %
\includegraphics[width=0.30\textwidth]{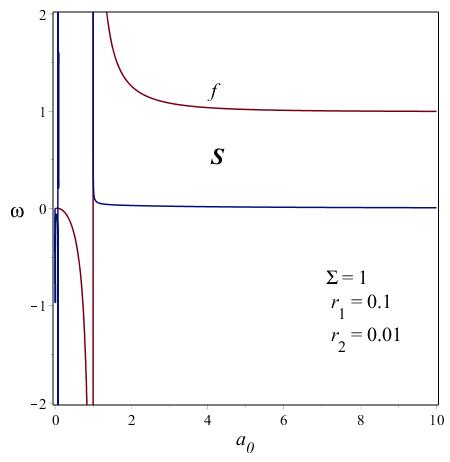} %
\includegraphics[width=0.30\textwidth]{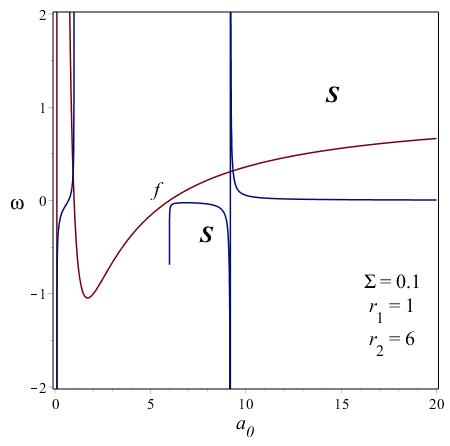} %
\caption{\small \textit{Stability regions of DTSW  in terms of $\omega$ and radius of the throat $a_{0}$ for different values of $\Sigma$, $r_{1}$ and $r_{2}$.} }
\end{figure}

\bigskip
\section{Conclusion}

In this work, we have constructed a stable DTSW in the context of EMD theory. In particular we explore the role of three parameters  $\Sigma$, $r_{1}$ and $r_{2}$, which encodes the effects of electric charge $Q$, magnetic charge $P$, and dilaton charge $\phi_{0}$, on the wormhole stability.  The surface stress at the
wormhole throat are computed via Darmois-Israel formalism while the stability analyses is carried out by using three different models. As a first model, we consider a LBG and show that the wormhole can be stable by choosing suitable values of parameters $\Sigma$, $r_{1}$ and $r_{2}$. In the second case, we focus on the stability analyses using a CG for the exotic matter and show that by choosing suitable values of parameters $\Sigma$, $r_{1}$ and $r_{2}$. Finally we use LogG for the exotic matter and show similar results. The results show that electric charge, magnetic charge, and dilaton charge play an important role on DTSW by increasing the stability domain of the wormhole. A particularly interesting finding is that for suitable values of $a_{0}$, the stable solutions exist in the DTSW for each values of $\omega$ that are chosen. We conclude that DTSW is linearly stable for variable EoS which supports the fact that the presence of EoS and dilaton/ electric/ magnetic charges of  induces stability in the WH geometry. In particular, we can see from the first two plots in Figure 1 that by keeping $r_1$ and $r_2$ fixed, and changing $\Sigma$, the vertical asymptote is slightly shifted to the left with the increase of $\Sigma$. This clearly indicates the increase of the stability region in the right of the asymptote. On the other hand, another model of charged TSW were constructed by Eiroa and Simeone \cite{eiroa2} in low energy string gravity which support our results. Their results show that the $\rho$ and $p$ on the shell for the null dilaton coupling parameter and it is showed that exotic matter is localized, moreover they managed to minimize the exotic matter needed using the  stronger dilaton-Maxwell coupling.

\section*{Competing Interests}
The authors declares that there is no conflict of interest regarding the publication of this paper.
\begin{acknowledgments}

This work was supported by the Chilean FONDECYT Grant No. 3170035 (A\"{O}). The authors thank to referees for their valuable suggestions.
\end{acknowledgments}

\end{document}